\input epsf
\documentstyle[12pt]{article}
\catcode`\@=11
\@addtoreset{equation}{section}

\catcode`\@=12

\title{What Schr\"{o}dinger's Cat is Telling}
\author{Guang-jiong NI\thanks{Email address: gjni@srcap.stc.sh.cn}\\
Department of Physics, Fudan University, Shanghai, 200433, China}     
\begin{document}
\date{}
\maketitle
\begin{abstract}
The recent experiment on superconducting quantum interference device provides new opportunity to clarify the fundamental interpretation of quantum mechanics. After analyzing this and relevant experiments, we claim the point of view that the quantum state in abstract sense contains no information and the wave function is the probability amplitude of fictitious measurement corresponding to the real one which is prepared to be done by the observer. Talking about the quantum state or wave function too materialized in ordinary language would often cause misunderstanding. So basically, the Schr\"{o}dinger's cat paradox is over.
\vskip 5mm

\end{abstract}
\vskip 1cm
\baselineskip=7mm

Recently, a beautiful experiment by Friedman et al provided strong evidence that a superconducting quantum interference device (SQUID) can be put into a superposition of two magnetic-flux states: one corresponding to a few microamperes of current flowing clockwise, the other corresponding to the same amount of current flowing anticlockwise[1].

As further explained by Blatter [2], the above experiment for the first time realized the thought experiment of Schr\"{o}dinger's dead-and-alive cat [3] in a truly macroscopic scale. So it has been attracting serious attention not only in science community, but also in the public (see e.g. [4,5]). In this paper we would like to discuss the implication of the experiment further, i.e., try to understand what Schr\"{o}dinger's cat is telling us?

Another very important experiment relevant to this one was performed by Myatt et al [6]. They clearly demonstrated that ``how Schr\"{o}dinger's cat is killed'' due to the decoherence effect via the coupling between the system (cat) and its environment as speculated since 1980s [7] (see also [2]). These two experiments, [1] and [6], have pushed the kernel of ``cat'' puzzle---the quantum coherence or the linear superposition of distinct quantum states---in front of all of us in a very acute way.

In Ref.[1], the ``cat'' state $|\psi>$  is expressed as the superposition of two different fluxoid states, $|0>$  and $|1>$, carrying clockwise and anticlockwise currents respectively:
$$|\psi >=C_0|0>+C_1|1> \eqno{(1)}$$                                             
with normalization condition $|C_0|^2+|C_1|^2=1$ and energy $E_{\psi}$. As stressed in Ref.[1] that the states $|0>$  and $|1>$  differ in flux by more than $1/4$ (flux quantum) and differ in current by $2-3\; \mu A$ which corresponds to the motion of $10^9$ electron (Cooper) pairs and a magnetic moment of SQUID being about $10^{10} \mu _B$. In the energy-flux diagram (Fig. 1 in [1]) it seems that $|0>$  and  $|1>$ are truly distinct macroscopic states separated by energy barrier with height $\Delta U_0$ and coupled together via the quantum tunnelling effect but located basically at the left and right potential wells respectively. Friedman et al prepared the system initially in the lowest state of left well (labeled $|i>$  with energy $E_i$) and applied pulse of microwave radiation (with photon energy $h\nu=96\; GHz=4.61K$) to excite the system into one $|\psi >$  state shown as (1) once when the energy condition ($E_{\psi}=E_i+h\nu$) is matched. Subsequently, the $|\psi >$  state decays into the right well. This transition between wells results in a change in flux and can be detected by the magnetometer. For a fixed $\Delta U_0$, the transition probability $P$ measured is proportional to $|C_0|^2$. The data of P as a function of external flux $\Phi_x$ (applied to SQUID loop for tilting the potential well) show a two-peak pattern which in turn reveals the enegy-level anticrossing between two states, say, $|\psi_+>=\frac{1}{\sqrt{2}}(|0>+|1>)$  and  $|\psi_->=\frac{1}{\sqrt{2}}(|0>-|1>)$     (special cases of $|\psi >$ , e.g., when $\Delta U_0=8.956K$). Thus the existence of the macroscopic ``cat'' state is verified.
As explained in [1], all the experimental data are consistent and in good agreement with the theoretical calculation by quantum mechanics (QM). What we will discuss is about some language we used above, i.e., some understanding about QM. Why we were so annoyed for the Schr\"{o}dinger's cat paradox? It is because so distinct two states, $|0>$  and $|1>$  , (i.e., ``dead cat'' and ``alive cat'') are superposed linearly (coherently) each other as shown in Eq.(1). In terms of wavefunction (WF) by its simulation of one particle in one-dimensional space as in Ref.[2], it reads:
$$\psi >(x,t)=<x|\psi (t)>=C_0\psi_0(x,t)+C_1\psi_1(x,t). \eqno{(2)}$$          
Here the WF in configuration space, i.e., in ``$x$ representation'' is defined as the contraction (projection) of state $|\psi (t)$  (``ket'' vector in Dirac's notation) with $<x|$  [``bra'' vector, the Hermitian conjugate of $|x>$ , which is the eigenstate (ket vector) of position operator $\hat{x}]$. The WF in QM is always a complex variable function, e.g., for a stationary state with energy $E$, $\psi (x,t)=\psi (x)\exp (-iEt/ \hbar )$. So WF is always non-observable. But what is observable? Max Born taught us that the modulus square of WF, i.e.,
$$|\psi >(x,t)|^2=|C_0|^2|\psi_0 (x,t)|^2+|C_1|^2|\psi_1 (x,t)|^2+2\mbox{Re}[C_0^*C_1\psi^*(x,t)\psi_1(x,t)]\eqno{(3)}$$ 
is proportional to the ``probability'' of finding the particle at point $x$ ``during'' the measurement. First, we wish to stress two words: ``probability'' rather than ``certainty'', ``during'' instead of ``before''. It seems to us that while the former caused no confusion, the latter was often overlooked in most textbooks on QM. Next, the third term in the right side of Eq.(3) is called the interference term. Though it's very small in this two-well problem due to tiny overlapping of $\psi_0(x)$  with $\psi_1(x)$  , it is usually quite important in other problems. The presence of interference term even denies the similar explanation for $|\psi_0(x)|^2$   and $|\psi_1(x)|^2$   separately. Only after we integrate Eq.(3) in the whole space, can we get the normalization condition:
$$\int |\psi(x)|^2dx=|C_0|^2+|C_1|^2=1\eqno{(4)}$$ 
with the interference term vanishing due to the orthogonality condition between $\psi_0(x)$  and $\psi_1(x)$  in general. Eq.(4) allows a simple probability interpretation and it is just the $|C_0|^2$ which was measured via the absorption of microwave in the experiment [1].

Of course, we can say that in the SQUID experiment[1] $|C_0|^2$ is the probability of finding the fluxoid state carrying clockwise current during the measurement. But we can't say so before the measurement since there is no $|C_0|^2$ in the WF as shown in Eq.(2). This is because WF is not a probability but a "probability amplitude" (see Eq.(3)).

Nowadays, nearly all physicists agree that the measurement is bound to change the object by destroying the quantum coherence originally existing in the object (or between it and its environment). Sometimes, it is said that the wave packet is collapsed in the measurement. However, we would like to add that there is no any information exists in the quantum state before the measurement. The information is created during the measurement by subject (the observer via the apparatus) and object in common. For example, the $|\psi_+>$  state of SQUID in symmetric well (the tilt $\varepsilon =0$) case is a stationary state carrying neither flux nor current. When $|C_0|>|C_1|$, state  $|0>$ dominates  $|1>$, only clockwise current could be observed in all experiments before the experiment[1], implying that all electrons move along one direction. This is because $|1>$  being in a subordinate position, one could at most say that "the anticlockwise current state is hidden in some sense." The situation will be in the reverse when $|C_0|<|C_1|$. So when we talk about $|\psi>$  being composed of   $|0>$ and $|1>$   carrying opposite currents and each with probability $|C_0|^2$ and $|C_1|^2$ respectively, we have to be cautious. We are already talking about its description, i.e., at the level of WF (not merely at the level of state in abstract sense). This is because the authors in [1] are going to measure the state via its interwell transition by magnetic flux change experiment, they need to resort to expansion (2) for predicting in advance what will happen statistically in a real experiment. Then as expected, the analysis of experimental data shows the component of   $|0>$ state in $\psi>$, i.e., $|C_0|^2$ contributes to the probability of making an interwell transition. Once a transition does occur in the measurement, the state $\psi>$  as a whole (rather than part of it) is gone.
Hence, we interpret WF as the probability amplitude of a fictitious
measurement. The word ``fictitious'' means that the measurement is not a
real one so it does not destroy the quantum coherence. We can evaluate
WF theoretically to predict what will happen statistically in a real
experiment ([8], see also [9]).

Actually, different points of view are subjected to even more severe test in a remarkable ``Which-Way(WW)'' experiment by D\"{u}rr, Nonn and Rempe [10](DNR). In their atom interferometer, in order to determine which one of the double-slit (which is realized by the Bragg diffraction of atom beam on two standing light waves) the $^{85}$Rb atoms go through, two microwave pulses are applied to designate the internal states $|2>$  and $|3>$  of atoms, one for each slit. Once this is done, the interference fringes are smeared due to the orthogonality condition $ <2|3>=0$. The analysis by DNR shows that the momentum transfer ($\Delta p$) from standing light wave or microwave pulse to atom is irrelevant or too tiny to account for the loss of interference. So DNR and some other physicists were considering the possibility that complementarity may not be enforced by the uncertainty relation:
$$\Delta p\Delta x\ge\hbar/2.\eqno{(4)}$$ 
In other words, the validity or generality of uncertainty principle is challenged.

In Ref.[8] we tried to understand the implication of DNR experiment via the examination on the essence of measurement. Since the WW information is got from the internal state instead of the impact of photon on the atom, the quantum coherence of WF of atom's center-of-mass motion has not been destroyed. As no momentum transfer is measured, the information about momentum $p$ itself does not exist at all, so does the relation (5). We need not worry about something which has not emerged yet. In fact, as a substitution of Eq.(5), DNR experiment has its own uncertainty relation that the distinguishability (of WW information) D and the fringe vsibility V are limited by the duality relation $D^2+V^2<1$ [10].

Therefore, in our point of view, the physics experiments accomplished
so far (especially in recent years, some of them were discussed in
Ref.[8]) all strongly support the validity of QM, which stands even
more firm than ever before. Basically, the "cat" paradox is
over. Indeed, schr\"{o}dinger's cat is now not only fat, but also
telling us something which we might not be aware of in the past. We
were often talking about the quantum state or WF too materialized in
ordinary language which caused misunderstanding. A state before
measurement is a linear (coherent) superposition of substates which
carry no information. Further probability interpretation of WF on the
state is only meaningful when it is related to certain measurement
which is prepared to be done. Another kind of measurement on the same
state may need another expansion in substates and corresponding
different probability interpretation. The ``fat cat'' is man-made, she
was not there originally. In contrast to some authors claim [11], we
believe that the quantum theory needs ``interpretation'' as discussed in
Ref [12]. And the correct interpretation of QM does need some
abstractness, more strictness and more flexibility as well. Physics is
now linked to philosophy more intimately than it used to be
[13,14,15], see Appendix.
\vskip 4mm

\section*{Appendix. What is the ``physical reality'' and how QM works?}

Nature stands itself independent of mankinds' consciousness. An object
is something absolute, infinite and containing no information (``thing
in itself''). But for cognizing it, we divide an individual from the
whole, i.e., separate a system (quantum state) from its environment
approximately by certain boundary condition. Only after some
measurement performed by us on it does an observed phenomenon (``thing
for us'') with some information occur. Being the probability amplitude
of relevant fictitious measurement, the wave function in corresponding
representation can predict in advance what will happen statistically
if the measurement is really made. It is just the power of QM. See
Fig. 1.

\begin{figure}[htbp]
  \begin{center}
\centerline{\epsfxsize=15cm\epsfbox{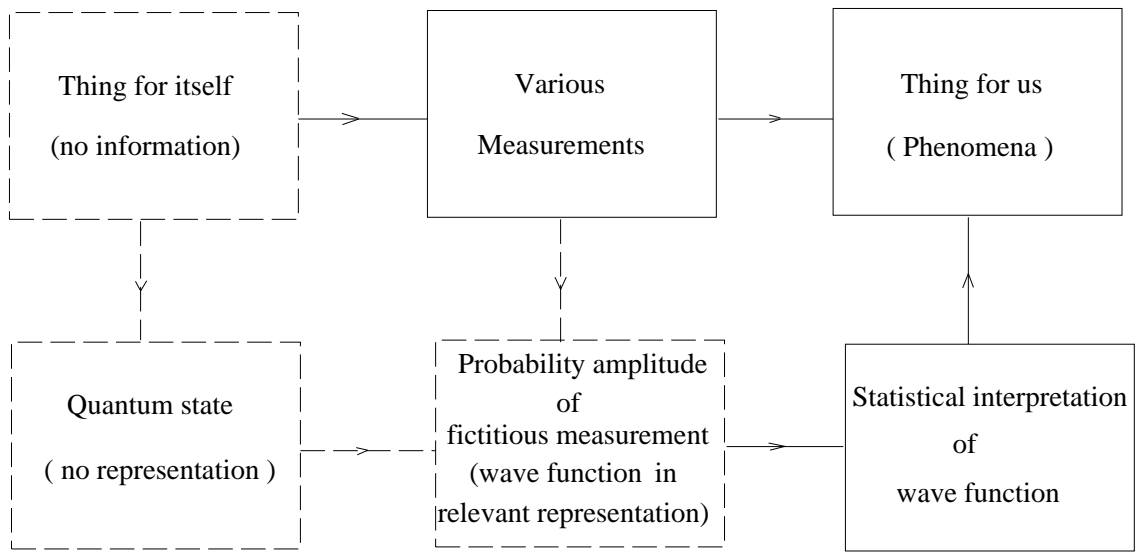}}    
    \caption{``physical reality'' should be defined at two levels linked by measurement and QM works in parallel.}
    \label{fig:fig.1}
  \end{center}
\end{figure}

\section*{Acknowledgement}
The author wishes to thank Dr. Z.-Y. Shen for bringing the
Refs. [1,2,4,5,11,12] to his attention as quickly as possible and relevant discussions.


\begin{thebibliography}{99}
\bibitem{15} [Note] We would like to say more about a "principle of relativity" in epistemology on which all of our discussions are based. It was stressed by Einstein when he established the theory of special relativity. He taught us that one should not discuss the absolute motion, absolute space or absolute time. Rather, one should discuss the relative motion, relative space or relative time. In general, a thing can only be cognized during its motion and change relative to other things. If isolated from its opposite, it would be bound to become mysterious object devoid of any under -standing. This is why for getting some description on the abstract state vector $|\psi (t)>$, one needs an another base vector $|x>$  to introduce the WF $\psi (x,t)=<x|\psi (t)>$. For understanding the existence state of individual (from a particle to Schr\"{o}dinger's cat, even to a human being), we need to survey its environment.

Interesting enough, similar principle had been comprehended by a young man, Jia Bao-yu, in the Chinese classic novel titled "Dream of the Red Mansion". He wrote as follows:

You won't be you if I wasn't born,

One can't understand her from her alone.
\end{thebibliography}
\end{document}